\documentclass[runningheads]{llncs}
\pdfoutput=1
\usepackage[mathletters]{ucs}
\usepackage[utf8x]{inputenc}
\usepackage[english]{babel}
\usepackage{microtype}
\usepackage{todonotes}
\usepackage[hidelinks]{hyperref}
\usepackage{amsmath}
\usepackage{cleveref}
\usepackage{listings}
\usepackage{pgfplots}
\usepackage{graphicx}
\usepackage{subfigure}
\usepackage{booktabs}
\usepackage[noend]{algpseudocode}
\usepackage{algorithm}

\title{Online Machine Learning Techniques for Coq: \\
A Comparison\thanks{
This work was supported
by the ERC grant no.\ 714034 \textit{SMART},
by the European Regional
Development Fund under the project AI\&Reasoning (reg. no.
CZ.02.1.01/0.0/0.0/15\_003/0000466),
and by the Ministry of Education, Youth and Sports within the dedicated program ERC CZ under the project POSTMAN no.~LL1902.
    }}

\author{
Liao Zhang\inst{1,3},
Lasse Blaauwbroek\inst{1,2},
Bartosz Piotrowski\inst{1,4},
Prokop Černý\inst{1},
Cezary Kaliszyk\inst{3,4}
and
Josef Urban\inst{1}
}

\authorrunning{L. Zhang et al.}
\institute{
Czech Technical University, Prague, Czech Republic
\and
Radboud University, Nijmegen, The Netherlands
\and
University of Innsbruck, Austria
\and
University of Warsaw, Poland
}

\MakeRobust{\Call}
\makeatletter
\renewcommand\section{\@startsection{section}{1}{\z@}%
                       {-12\p@ \@plus -4\p@ \@minus -4\p@}%
                       {8\p@ \@plus 4\p@ \@minus 4\p@}%
                       {\normalfont\large\bfseries\boldmath
                        \rightskip=\z@ \@plus 8em\pretolerance=10000 }}
\makeatother

\begin{document}
\maketitle % typeset the header of the contribution
\begin{abstract}
  We present a comparison of several online machine learning techniques for
  tactical learning and proving in the Coq proof assistant. This work builds on
  top of Tactician, a plugin for Coq that learns from proofs written by the user
  to synthesize new proofs. Learning
  happens in an online manner, meaning that Tactician's machine learning model
  is updated immediately every time the user performs a step in an interactive
  proof. This has important advantages compared to the more studied
  offline
  learning systems: (1) it
  provides the user with a seamless, interactive experience with Tactician and,
  (2) it takes advantage of locality of proof similarity, which means that proofs similar
  to the  current proof are likely to be found close by.
  We implement two online methods,
  namely approximate $k$-nearest neighbors based on locality sensitive
  hashing forests and random decision forests. Additionally,
  we conduct experiments with gradient boosted trees in an offline setting
  using XGBoost.
  We compare the relative performance of Tactician using these three
  learning methods on Coq's standard library.

\end{abstract}

\section{Introduction}
\label{sec:introduction}

The users of interactive theorem proving systems are in dire need of a digital sidekick,
which helps them reduce the time spent proving the mundane parts of their
theories, cutting down on the man-hours needed to turn an informal theory into a
formal one. The obvious way of creating such a digital assistant is using
machine learning. However, creating a practically usable assistant comes with
some requirements that are not necessarily conducive to the most trendy machine learning
techniques, such as deep learning.

The environment provided by ITPs is highly
dynamic, as it maintains an ever-changing global context of definitions, lemmas,
and custom tactics. Hence, proving lemmas within such environments requires intimate knowledge
of all the defined objects within the global context. This is contrasted by---for
example---the game of chess; even though the search space is enormous,
the pieces always move according to the same rules, and no new kinds of pieces can
be added. Additionally, the interactive nature of ITPs
demands that machine learning techniques do not need absurd amounts of time and resources to
train (unless a pre-trained model is highly generic and widely applicable
across domains; something that has not been achieved yet). In this paper, we are
interested in online learning techniques that quickly learn from
user input and immediately utilize this information. We do this in the context
of the Coq proof assistant~\cite{the_coq_development_team_2019} and specifically
Tactician~\cite{DBLP:conf/mkm/BlaauwbroekUG20}---a plugin for
Coq that is designed to learn from the proofs written by a user and apply that
knowledge to prove new lemmas.

Tactician performs a number of functions, such as proof recording, tactic
prediction, proof search, and proof reconstruction. In this paper, we focus on
tactic prediction. For this, we need a machine learning technique that accepts as
input a database of proofs, represented as pairs containing a proof state and the tactic
that was used to advance
the proof. From this database, a machine learning model
is built. The machine learning task
is to predict an appropriate tactic when given a proof state. Because the model needs to
operate in an interactive environment, we pose four requirements the learning
technique needs to satisfy:
\begin{enumerate}
\item The model (datastructure) needs to support dynamic updates. That is, the
  addition of a new pair of a proof state and tactic to the current model needs
  to be done in (near) constant time.
\item The model should limit its memory usage to fit in a consumer laptop. We
  have used the arbitrary limit of 4 GB.
\item The model should support querying in (near) constant time.
\item The model should be persistent (in the functional programming
  sense~\cite{DBLP:journals/jcss/DriscollSST89}). This enables the model to be synchronized with the
  interactive Coq document, in which the user can navigate back and forth.
\end{enumerate}

\subsection{Contributions}
In this work, we have implemented two online learning models. An improved version of the locality
sensitive hashing scheme for $k$-nearest neighbors is described in detail in
\Cref{sec:lshf}. An implementation of random forest is described in
\Cref{sec:random-forest}. In \Cref{sec:eval}, we evaluate both models, comparing
the number of lemmas of Coq's standard library they can prove in a
chronological setting (i.e., emulating the growing library).

In addition to the online models, as a proof of concept, we also
experiment in an offline fashion with boosted trees, specifically
XGBoost~\cite{chen2016xgboost} in \Cref{sec:boosted}. Even though the model learned by XGBoost cannot be used
directly in the online setting described above, boosted trees are today among
the strongest learning methods. Online algorithms for
boosted trees do exist~\cite{DBLP:journals/npl/ZhangZSAFS19}, and we intend to implement them in the future.

The techniques described here require representing proof states as feature
vectors. Tactician already supported proof state representation using simple
hand-rolled features~\cite{blaauwbroek2020tactic}. In addition, \Cref{sec:features} describes
our addition of more advanced features of the proof states, which are shown to improve Tactician's performance in \Cref{sec:eval}.

\section{Tactic and Proof State Representation}
\label{sec:features}
To build a learning model, we need to characterize proof states and the tactics applied to them.
To represent tactics, we first perform basic decompositions and simplifications and denote the resulting atomic tactics by their hashes \cite{blaauwbroek2020tactic}.

Tactician's original proof state features~\cite{blaauwbroek2020tactic} consist
merely of identifiers and adjacent identifier pairs in the abstract syntax tree (AST).
Various other, more advanced features have been considered for automated reasoning systems built over large formal mathematical knowledge bases~\cite{ChvalovskyJ0U19,tgckju-lpar17,kaliszyk2015efficient}.
To enhance the performance of Tactician, we modify the old feature set and define new features as follows.

\paragraph{\textbf{Top-down Oriented AST Walks}} We add top-down oriented walks in the AST of length up to 3 with syntax placeholders. For instance, the unit clause $f(g(x))$ will contain the features:
	\begin{lstlisting}[basicstyle=\ttfamily]
f:AppFun, g:AppFun, x:AppArg, f:AppFun(g:AppFun),
g:AppFun(x:AppArg), f:AppFun(g:AppFun(x:AppArg))
	\end{lstlisting}
	The feature \texttt{g:AppFun} indicates that $g$ is able to act as a function in the term tree, and \texttt{x:AppArg} means that $x$
        is only an argument of a function.

\paragraph{\textbf{Vertical Abstracted Walks}} We add  vertical walks in the term tree from the root to atoms in which nonatomic nodes are substituted by their syntax roles. For the term  $f_1(f_2(f_3(a)))$, we can convert each function symbol to \texttt{AppFun} whereas the atom $a$ is transformed to \texttt{a:AppArg} as above.
	Subsequently, we can export this as the feature \texttt{AppFun(AppFun(AppFun(a:AppArg)))}.
	Such abstracted features are designed to better capture the overall abstract structure of the AST.

\paragraph{\textbf{Top-level Structures}} We add top-level patterns by replacing
the atomic nodes and substructures deeper than level 2 with a single symbol \texttt{X}.
	Additionally, to separate the function body and arguments, we append the arity of the function to the corresponding converted symbol.
	As an example, consider the term $f(g(b,c),a)$ consisting of atoms $a,b,c,f,g$. We first replace $a,f,g$ with \texttt{X} because they are atomic.
	We further transform $f$ and $g$ to \texttt{X2} according to the number of their arguments.
	However, $b$ and $c$ break the depth constraint and should be merged to a single \texttt{X}.
	Finally, the concrete term is converted to an abstract structure \texttt{X2(X2(X),X)}.
	Abstracting a term to its top-level structure is useful for
	determining whether a ``logical" tactic should be applied. As an illustration, the presence of $X \land X$ in the goal often indicates that we should perform case analysis by the \texttt{split} tactic. Since we typically do not need all the nodes of a term to decide such structural information, and we want to balance the generalization with specificity, we use the maximum depth 2.

\paragraph{\textbf{Premise and Goal Separation}} Because local hypotheses
typically play a very different role than the conclusion of a proof state, we separate their feature spaces. This can be done by serially numbering the features and adding a sufficiently large constant to the goal features.

\paragraph{\textbf{Adding Occurrence Counts}} In the first version of Tactician, we have used only a simple boolean version of the features. We try to improve on this by adding the number of occurrences of each feature in the proof state.

\section{Prediction Models}
\label{sec:models}

\subsection{Locality Sensitive Hashing Forests for Online $k$-NN}
\label{sec:lshf}

One of the simplest methods to find correlations between proof states is to
define a metric or similarity function $d(x, y)$ on the proof states. One can
then extract an ordered list of length $k$ from a database of proof states that are as
similar as possible to the reference proof state according to $d$. Assuming that
$d$ does a good job identifying similar proof states, one can then use tactics
known to be useful in a known proof state for an unseen proof state. In
this paper, we refer to this technique as the $k$-nearest neighbor ($k$-NN) method (even
though this terminology is somewhat overloaded in the literature).

Our distance function is based on the features described in \Cref{sec:features}.
We compare these features using the Jaccard index $\text{J}(f_1, f_2)$.
Optionally, features can be weighted using the TfIdf statistic~\cite{DBLP:journals/jd/Jones04}, in
which case the generalized index $\text{J}_w(f_1, f_2)$ is used.
\[ \text{J}(f_1, f_2) = \frac{|f_1\cap f_2|}{|f_1\cup f_2|} \quad
\text{tfidf}(x) = \log \frac{N}{|x|_N} \quad \text{J}_w(f_1, f_2) =
\frac{\sum_{x\in f_1\cap f_2} \text{tfidf}(x)}{\sum_{x\in f_1\cup f_2} \text{tfidf}(x)}\]
Here $N$ is the database size, and $|x|_N$ is the number of times feature $x$
occurs in the database. In previous work, we have made a more detailed comparison of similarity functions~\cite{blaauwbroek2020tactic}.

A naive implementation of the $k$-NN method is not very useful in
the online setting because the time complexity for a query grows linearly with
the size of the database. Indexing methods, such as k-d trees, exist to speed
up queries~\cite{DBLP:journals/cacm/Bentley75}. However, these methods do not scale well when the
dimensionality of the data increases~\cite{DBLP:journals/toc/Har-PeledIM12}. In this work, we instead implement an approximate
version of the $k$-NN method based on Locality Sensitive Hashing (LSH)~\cite{DBLP:conf/vldb/GionisIM99}.
This is an upgrade of our previous LSH implementation that was not persistent and was slower.
We also describe our functional implementation of the method in detail for the first time here.

The
essential idea of this technique is to hash feature vectors into buckets using a
family of hash functions that guarantee that similar vectors hash to the same bucket with
high probability (according to the given similarity function). To find a $k$-NN
approximation, one can simply return the contents of the bucket
corresponding to the current proof state. For the Jaccard index, the appropriate
family of hash functions are the MinHash functions~\cite{DBLP:conf/sequences/Broder97}.

The downside of the naive LSH method is that its
parameters are difficult to tune. The probability that the vectors that hash to
the same bucket are similar can be increased by associating more than one hash
function to the bucket.
All values of the hash functions then need to pair-wise agree for the
items in the bucket. However, this will naturally decrease the size of the
bucket, lowering the number of examples $k$ (of $k$-NN) that can be retrieved.
The parameter $k$ can be increased again by simply maintaining multiple independent
bucketing datastructures. Tuning these parameters is critically dependent on the
size of the database, the length of the feature vectors, and the desired value of
$k$. To overcome this, we implement a highly efficient, persistent, functional variant of Locality
Sensitive Hashing Forest~\cite{DBLP:conf/www/BawaCG05} (LSHF), which is able to tune these
parameters automatically, leaving (almost) no parameters to be tuned manually.
Below we give a high-level overview of the algorithm as it is modified for a
functional setting. For a more in-depth discussion on the correctness of the
algorithm, we refer to the previous reference.

LSHFs consist of a forest (collection) of tries
$\mathcal{T}_1\ldots\mathcal{T}_n$. Every trie has an associated hash function
$h_i$ that is a member of a (near) universal hashing family mapping a feature
down to a single bit (a hash function mapping to an integer can be used by
taking the result modulus two). To add a new
example to this model, it is inserted into each trie according to a path (sequence) of
bits.
Every bit of this path can be shown to be locally sensitive for the Jaccard
index~\cite{DBLP:conf/www/BawaCG05}. The path
of an example is calculated using the set of features that represents the proof
state in the example.
\[\text{path}_i(f) = \text{sort}(\{h_i(x) \mid x \in f \})\]
For a given trie $\mathcal{T}$, the subtrie starting at a given path
$b_1\ldots b_m$ can be seen as the bucket to which examples that agree on the
hashes $b_1\ldots b_m$ are assigned. Longer paths point
to smaller buckets containing less similar examples, while shorter paths point to
larger buckets containing increasingly similar examples. Hence, to retrieve the neighbors of a proof state
with features $f$, one should start by finding examples that share the entire
path of $f$. To retrieve more examples, one starts collecting the subtrees
starting at smaller and smaller prefixes of $\text{path}_i(f)$. To increase the
accuracy and number of examples retrieved, this procedure can be performed on
multiple tries simultaneously, as outlined in Algorithm $\ref{alg:lshf}$.

\begin{algorithm}[tb!]
\caption{Querying the Locality Sensitive Hashing Forest}\label{alg:lshf}
\begin{algorithmic}[1]
  \Function {QueryLSHF}{$\mathcal{F}$, $f$}\Comment{$\mathcal{F}$ a forest, $f$
    a feature set}
  \State $\mathcal{P}\gets \langle \text{path}_i(f) : i \in [1..|\mathcal{F}|]
  \rangle$
  \State neighbors $\gets$ \Call{FilterDuplicates}{\Call{SimultaneousDescend}{$\mathcal{F}$, $\mathcal{P}$}}
  \State Optionally re-sort neighbors according to real Jaccard index
  \EndFunction
  \Function{SimultaneousDescend}{$\mathcal{F}$, $\mathcal{P}$}
  \State $\mathcal{F}_\text{rel}\gets\langle$ if head($\mathcal{P}$) then
  left($\mathcal{T}$) else right($\mathcal{T}$) $: \mathcal{T} \in \mathcal{F}$ when not leaf($\mathcal{T}$) $\rangle$
  \State $\mathcal{F}_\text{irrel}\gets\langle$ if leaf($\mathcal{T}$) then
  $\mathcal{T}$ elseif head($\mathcal{P}$) then
  right($\mathcal{T}$) else left($\mathcal{T}$) $: \mathcal{T} \in \mathcal{F}\ \rangle$
  \If {$\mathcal{F}_\text{rel}$ is empty}
  \State neighbors $\gets$ empty list
  \Else
  \State $\mathcal{P}' \gets \langle \text{tail}(\mathcal{P}_i) : i \in [1..n] \rangle$
  \State neighbors $\gets$ \Call{SimultaneousDescend}{$\mathcal{F}_\text{rel}$, $\mathcal{P}'$}
  \EndIf
  \If {$|\text{neighbors}| \geq k$}
  \State\Return neighbors
  \Else
  \State\Return \Call{Append}{neighbors, \Call{Concatenate}{$\langle$
    \Call{Collect}{$\mathcal{T} : \mathcal{T} \in \mathcal{F}_\text{irrel}\rangle$}}}
  \EndIf
  \EndFunction
\end{algorithmic}
\end{algorithm}

Tuning the LSHF model consists mainly of choosing the appropriate number of
tries that maximizes the speed versus accuracy trade-off. Experiments show that
$11$ trees is the optimal value. Additionally, for efficiency reasons, it is a
good idea to set a limit on the depth of the tries to prevent highly similar
examples from creating a deep trie. For our dataset, a maximum depth of $20$ is
sufficient.

\subsection{Online Random Forest}
\label{sec:random-forest}
Random forests are a popular machine learning method combining many randomized
decision trees into one ensemble, which produces predictions via voting~\cite{rf}.
Even though the decision trees are not strong learners on their own, because they
are intentionally decorrelated, the voting procedure greatly improves on top of their
individual predictive performance. The decision trees
consist of
internal nodes labeled by decision rules and leaves labeled by
examples. In our case, these are tactics to be applied in the proofs.

Random forests are a versatile method that requires little tuning of
its hyperparameters. Their architecture is also relatively simple, which
makes it easy to provide a custom OCaml implementation easily integrable
with Tactician, adhering to its requirement of avoiding mutable data
structures.
Direct usage of existing random forest implementations is impossible due to
challenges in Tactician's learning setting. These
challenges are: (1) numerous sparse features, (2) the necessity of online
learning, as detailed in the next two paragraphs.

The decision rules in nodes of the decision trees are based on the features
of the training examples. These rules are chosen to maximize the
\textit{information gain}, i.e., to minimize the \textit{impurity} of the set of
labels in the node.\footnote{If we have labels $\{a, a, b, b, b\}$, ideally, we
would like to produce a split which passes all the examples with label $a$ to
one side and the examples with $b$ to the other side.} There are more than $37,000$
binary and sparse features in Tactician. Since
the learner integrated with Tactician needs to be fast, one needs to be careful
when optimizing the splits in the tree nodes.

Random forests are typically trained in an offline manner where the whole
training data is available at the beginning of the training. In Tactician this would be quite suboptimal. To take advantage of the locality of proof
similarity and to be able to use data coming from new proofs written by a user,
we want to immediately update the machine learning model
after each proof.

\begin{algorithm}[tb!]
\algsetblockdefx[match]{Match}{EndMatch}{}{}[1]{\textbf{match} #1 \textbf{with}}{aa}
\algsetblockdefx[with]{With}{EndWith}{}{}[1]{{#1}\textbf{:}}{aa}
\algnotext[with]{EndWith}
\algnotext[match]{EndMatch}
\caption{Adding training a example $e$ to a decision tree $\mathcal{T}$}\label{alg:tree}
\begin{algorithmic}[1]
\Function {AddExampleToTree}{$\mathcal{T}$, $e$}
\Match{$\mathcal{T}$}
    \With{Node($\mathcal{R}$, $\mathcal{T}_l$, $\mathcal{T}_r$)}
    \Comment{$\mathcal{R}$ -- binary rule, $\mathcal{T}_l$, $\mathcal{T}_r$ --
    left and right subtrees}
        \Match{$\mathcal{R}$($e$)}
            \With{Left}
            \Return Node($\mathcal{R}$, \textsc{AddExampleToTree}($\mathcal{T}_l$, $e$), $\mathcal{T}_r$)
            \EndWith
            \With{Right}
            \Return Node($\mathcal{R}$, $\mathcal{T}_l$, \textsc{AddExampleToTree}($\mathcal{T}_r$, $e$))
            \EndWith
        \EndMatch
    \EndWith
    \With{Leaf($l$, $\mathcal{E}$)}
    \Comment{$l$ -- label/tactic, $\mathcal{E}$ -- examples}
        \State $\mathcal{E}$ $\gets$ \textsc{Append}($\mathcal{E}$, $e$)
        \If {\textsc{SplitCondition}($\mathcal{E}$)}
            \State $\mathcal{R}$ $\gets$ \textsc{GenerateSplitRule}($\mathcal{E}$)
            \State $\mathcal{E}$$_l$, $\mathcal{E}$$_r$ $\gets$ \textsc{Split}($\mathcal{R}$, $\mathcal{E}$)
            \State $l_l$ $\gets$ label of random example from $\mathcal{E}_l$
            \State $l_r$ $\gets$ label of random example from $\mathcal{E}_r$
            \State \Return Node($\mathcal{R}$, Leaf($l_l$, $\mathcal{E}_l$),
            Leaf($l_r$, $\mathcal{E}_r$))
        \Else
            \State \Return Leaf($l$, $\mathcal{E}$)
        \EndIf
    \EndWith
\EndMatch
\EndFunction
\end{algorithmic}
\end{algorithm}

There are approaches to turn random forests into online learners
 \cite{orf,online_rf_2} which inspired our implementation.
The authors of \cite{orf}
propose a methodology where new training examples are passed to the leaves
of the decision trees, and under certain statistical conditions, the leaf is
split and converted to a new decision node followed by two new leaves.
We take a similar approach, but deciding a split in our implementation is
simpler and computationally cheaper.

The pseudocode describing our implementation is presented below. Algorithm~\ref{alg:tree} shows how the training examples are added to the decision trees.
A new training example is passed down the tree to one of its leaves. The
trajectory of this pass is governed by binary decision rules in the nodes of the
tree. Each rule
checks whether a given 
feature is
present in the example. Once the example reaches a leaf, it is saved there, and
a decision is made whether to extend the tree (using function
\textsc{SplitCondition}). This happens only when the Gini Impurity measure
\cite{ml} on the set 
of examples in the leaves is greater than a
given impurity threshold $i$ (a hyperparameter of the model). When the split is
done, the leaf becomes an internal node with a new split rule, and the
collected examples from the leaf are passed down to the two new leaves. The new rule
(an output from \textsc{GenerateSplitRule}) is produced in the following way.
$N$ features are selected from the features of the examples, where $N$
is the square root of the number of examples. The selection of the features is
randomized and made in such a way that features that are distinguishing
between the examples have higher probability: First, we randomly select two
examples from the leaf, and then we randomly select a feature from the difference of sets of features of the two examples.
Among such selected features, the one
maximizing the \textit{information gain} \cite{ml} of the split rule based on it is selected.
The two new leaves get labels randomly selected from the examples
belonging to the given leaf.

When adding an example to a random forest (Algorithm~\ref{alg:forest}), first,
a decision is made whether a new tree (in the form of a single leaf) should be added
to the forest. It happens with probability $\frac{1}{n}$, where $n$ is the number of
trees in the forest under the condition that $n$ is lower than a given threshold.

Predicting a tactic for a given example with a random forest (Algorithm~\ref{alg:pred}) is done in two steps. First, the example is passed to the leaves
of all the trees and the labels (tactics) in the leaves are saved. Then the ranking of
the tactics is made based on their frequencies.

\begin{algorithm}[tb!]
\caption{Adding a training example $e$ to a random forest $\mathcal{F}$}
\label{alg:forest}
\begin{algorithmic}[1]
\Function {AddExampleToForest}{$\mathcal{F}$, $e$, $n_\text{max}$}
\Comment{$n_\text{max}$ -- max number of trees}
\State $n$ $\gets$ number of trees in $\mathcal{F}$
\State $m$ $\gets$ random number from $\{1, \ldots n\}$
\State $\mathcal{F}_\text{updated}$ $\gets$ empty list
\If {n $<$ $n_\text{max}$ and $m = 1$}
\State $\mathcal{T} \gets$ leaf labeled by tactic used in $e$
\State $\mathcal{F}_\text{updated}$ $\gets$
\textsc{Append}($\mathcal{F}_\text{updated}$, $\mathcal{T}$)
\EndIf
\ForAll{$\mathcal{T} \in \mathcal{F}$}
\State $\mathcal{T} \gets$ \textsc{AddExampleToTree}($\mathcal{T}$, $e$)
\State $\mathcal{F}_\text{updated}$ $\gets$
\textsc{Append}($\mathcal{F}_\text{updated}$, $\mathcal{T}$)
\EndFor
\State \Return $\mathcal{F}_\text{updated}$
\EndFunction
\end{algorithmic}
\end{algorithm}

\begin{algorithm}[tb!]
\caption{Predicting labels for unlabeled $e$ in the random forest $\mathcal{F}$}
\label{alg:pred}
\begin{algorithmic}[1]
\algsetblockdefx[match]{Match}{EndMatch}{}{}[1]{\textbf{match} #1 \textbf{with}}{aa}
\algsetblockdefx[with]{With}{EndWith}{}{}[1]{{#1}\textbf{:}}{aa}
\algnotext[with]{EndWith}
\algnotext[match]{EndMatch}
\Function {PredictForest}{$\mathcal{F}$, $e$}
\State $\mathcal{P}$ $\gets$ empty list \Comment{$\mathcal{P}$ -- predictions}
\ForAll{$\mathcal{T}$ $\in$ $\mathcal{F}$}
\State $t$ $\gets$ \textsc{PredictTree}($e$)
\State append $t$ to $\mathcal{P}$
\EndFor
\State $R \gets$ \textsc{Vote}($\mathcal{P}$) \Comment{$R$ -- ranking of tactics}
\State \Return $R$
\EndFunction
\Function {PredictTree}{$\mathcal{T}$, $e$}
\Match{$\mathcal{T}$}
    \With {Node($\mathcal{R}$, $\mathcal{T}$$_l$, $\mathcal{T}$$_r$)}
        \Match{$\mathcal{R}(e)$}
            \With{Left}
                \Return \textsc{PredictTree}($\mathcal{T}$$_l$, $e$)
            \EndWith
            \With{Right}
                \Return \textsc{PredictTree}($\mathcal{T}$$_r$, $e$)
            \EndWith
        \EndMatch
    \EndWith
    \With{Leaf($l$, $\mathcal{E}$)}
        \Return $l$
    \EndWith
\EndMatch
\EndFunction
\end{algorithmic}
\end{algorithm}

\subsubsection{Tuning Hyperparameters}
\label{sec:rf-tuning}

\begin{figure}[tb!]
\caption{Results of hyperparameter tuning for random forests. The blue circle
corresponds to top-10 accuracy (how often the correct tactic was present in
the first 10 predictions) whereas the red square corresponds to top-1 accuracy.}
\label{fig:rf-tuning}
  \subfigure{
  \hspace{-0.3cm}
  \begin{minipage}[t]{0.50\textwidth}
	\begin{tikzpicture}
	\begin{axis}[
		xlabel={Number of trees}, ylabel={Accuracy}, width = 5.5cm, height = 4.5cm,
		xtick = {10, 20, 40, 80, 160, 320, 640},
		xticklabels = {10, 20, 40, 80, 160, 320, 640},
		xmode = log]
	\addplot table [x=n_trees, y=acc10, col sep=comma]{csv/n_trees.csv};
	\addplot table [x=n_trees, y=acc1, col sep=comma]{csv/n_trees.csv};
    \end{axis}
	\end{tikzpicture}
  \end{minipage}}
  \hspace{0.3cm}
  \subfigure{
  \begin{minipage}[t]{0.40\textwidth}
	\begin{tikzpicture}
	\begin{axis}[
		xlabel={Impurity threshold}, width = 5.5cm, height = 4.5cm,
		xtick = {0.1, 0.3, 0.5, 0.7, 0.9},
		xticklabels = {0.1, 0.3, 0.5, 0.7, 0.9},
        ]
	\addplot table [x=min_impur, y=acc10, col sep=comma]{csv/min_impur.csv};
	\addplot table [x=min_impur, y=acc1, col sep=comma]{csv/min_impur.csv};
    \end{axis}
	\end{tikzpicture}
  \end{minipage}}
\end{figure}
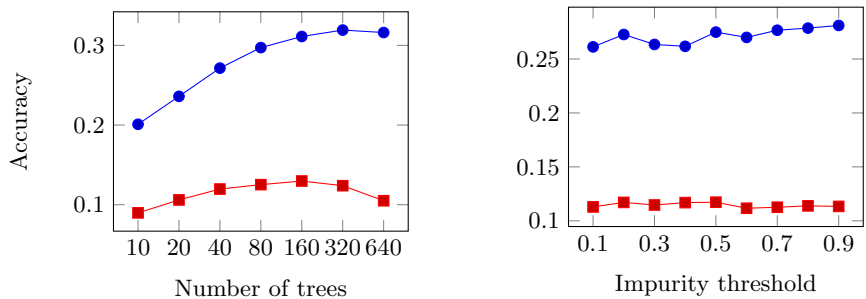

There are two hyperparameters in our implementation of random forests: (1)
the maximal number of trees in the forest and (2) the impurity threshold for 
performing the node splits. 
To determine the influence of
these parameters on the predictive power, we perform a grid search. For this,
we randomly split the data that is not held out for testing (see
Section~\ref{sec:split}) into a training and validation part. The results of the grid
search are shown in Figure~\ref{fig:rf-tuning}. The best numbers of trees are 160
(for top-1 accuracy) and 320 (for top-10 accuracy). We used these two values for
the rest of the experiments. For the impurity threshold, it is difficult to
see a visible trend in performance; thus we selected 0.5 as our default.

\subsection{Boosted Trees}
\label{sec:boosted}

Gradient boosted decision trees are a state-of-the-art machine learning
algorithm that transforms weak base learners, decision trees, into a method
with stronger predictive power by appropriate combinations of the base models. One
efficient and powerful implementation is the XGBoost
library. Here, we perform some initial experiments in an offline setting for
tactic prediction. Although XGBoost can at the moment not be directly integrated
with Tactician, this gives us a useful baseline based on existing
state-of-the-art technology.
Below, we illustrate a procedure of developing
our XGBoost model based on binary logistic regression.

The input to XGBoost is a sparse matrix containing rows with the format of
$(\phi_{P}, \phi_{T})$ where $\phi_{P}$ includes the features of a proof state,
and $\phi_{T}$ characterizes a tactic related to the proof state.
We transform each proof state to a sparse feature vector $\phi_{P}$ containing
the features' occurrence counts.
Since there may be a large number of features in a given Coq
development environment, which may hinder the
efficiency of training and prediction, it is reasonable to decrease the dimension of the vectors.
We hash the features to $20,000$ buckets by using the modulo of the feature's index.
As above, we also remap the tactic
hashes to a $20,000$-dimensional space separated from the state features.

The training examples get labels 1 or 0 based on the tactics being useful or not for the proof state.
A tactic for a certain proof state is labeled as positive if it is exactly the one applied
to this state in the library.
In contrast, negative tactics are elements in the tactic space that differ from
the positive instance. We obtain negative data by two
approaches: \textit{strong} negatives and \textit{random} negatives.
Strong negative instances are obtained by arbitrarily selecting a subset from the best-100 $k$-NN predictions for this state.
In the other approach, negative instances are arbitrarily
chosen from the entire tactic space.

With a trained gradient boosted trees model, we can predict the scores of the tactics for an unseen proof state $P$. First, the top-100 $k$-NN
predictions are preselected. Then, for each tactic, we input $(\phi_{P},
\phi_{T})$ to the model to obtain the score of $T$. The tactics are then sorted according to their scores.

\begin{figure}[tb!]
	\caption{Results of hyperparameter tuning for gradient boosted trees. In
		consistence with Figure~\ref{fig:rf-tuning}, the blue circle (red square) corresponds to top-10 (top-1) accuracy, respectively. The graph of negative ratios contains two additional curves of random negative examples. The brown circle relates to top-10 accuracy, whereas the black star presents the results of top-1 accuracy.}
	\label{fig:XGB-param-tuning}
	\centering
	\subfigure{
		\begin{minipage}[t]{0.48\textwidth}
			\centering
			\begin{tikzpicture}
				\begin{axis}[ xlabel={Negative ratios}, ylabel={Accuracy}, width = 5.5cm, xmode = log, log basis x={2},
					xtick = {1, 2, 4, 8,16, 32},
					xticklabels = {1, 2, 4, 8,16, 32},
					]
					\addplot table [x=neg-number, y=prob, col sep=comma, mark=*] {csv/xgb-neg-top10.csv};
					\addplot table [x=neg-number, y=prob, col sep=comma, mark=*] {csv/xgb-neg-top1.csv};
					\addplot table [x=neg-number, y=prob, col sep=comma,
					mark=*] {csv/xgb-rand-neg-top10.csv};
					\addplot table [x=neg-number, y=prob, col sep=comma,
					mark=*] {csv/xgb-rand-neg-top1.csv};
				\end{axis}
			\end{tikzpicture}
	\end{minipage}}
	\subfigure{
        \begin{minipage}[t]{0.48\textwidth}
			\centering
			\begin{tikzpicture}
				\begin{axis}[ xlabel={Number of trees}, width = 5.5cm,
					xmode = log, log basis x={2},
					xtick={1,8,64,512,4096},
					xticklabels={1,8,64,512,4096},
					]
					\addplot table [x=tree-number, y=prob, col sep=comma] {csv/xgb-tree-num-top10.csv};
					\addplot table [x=tree-number, y=prob, col sep=comma] {csv/xgb-tree-num-top1.csv};
				\end{axis}
			\end{tikzpicture}
	\end{minipage}}
	\subfigure{
		\begin{minipage}[t]{0.48\textwidth} \centering
			\begin{tikzpicture}
				\begin{axis}[
					xlabel={Eta parameters}, ylabel={Accuracy}, width = 5.5cm,
					xtick = {0.01, 0.04, 0.16, 0.64},
					xticklabels={0.01, 0.04, 0.16, 0.64},
					xmode = log, log basis x={2},
					]
					\addplot table [x=eta, y=prob, col sep=comma] {csv/xgb-eta-top10.csv};
					\addplot table [x=eta, y=prob, col sep=comma] {csv/xgb-eta-top1.csv};

				\end{axis}
			\end{tikzpicture}
	\end{minipage}}
	\subfigure{
		\begin{minipage}[t]{0.48\textwidth} \centering
			\begin{tikzpicture}
				\begin{axis}[ xlabel={Max depth of trees}, width = 5.5cm, xmode =
					log, log basis x={2},
					xtick = {1,2,4,8,16},
					xticklabels = {1,2,4,8,16},
					]
					\addplot table [x=depth, y=prob, col sep=comma] {csv/xgb-depth-top10.csv};
					\addplot table [x=depth, y=prob, col sep=comma] {csv/xgb-depth-top1.csv};
				\end{axis}
			\end{tikzpicture}
	\end{minipage}}
\end{figure}
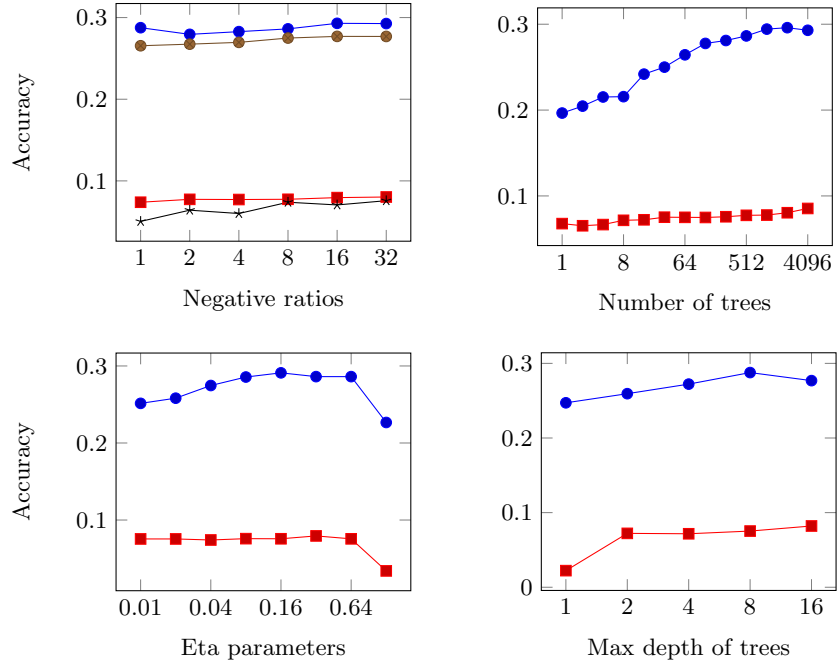

\subsubsection{Tuning Hyperparameters}
\label{sec:xgb-tuning}
Similarly as for the random forest model (Section~\ref{sec:rf-tuning}), we
optimize the most important hyperparameters of the XGBoost training algorithm on
the data coming from the non-sink nodes in the dependency graph of Coq's standard library
(see Section \ref{sec:split}).
One essential parameter is the \textit{ratio} of negative examples. Ratio $n$ indicates that we generate $n$ negative instances for each
recorded proof state.
Other influential parameters that we tune are: \textit{eta} (learning-rate), \textit{number of trees}, and \textit{max depth}.  Due to the limitations of computing resources, we assume a set of default parameters: $\textit{ratio}=8$, $eta=0.2$, $\textit{number of trees}= 500$, $\textit{max depth}=10$,  and then separately modify each of these parameters to observe the influence caused by the change, which is depicted in
Figure~\ref{fig:XGB-param-tuning}.
Both strong and random negatives are evaluated. 
Obviously, strong negatives perform better than random negatives, and increasing the negative ratios will certainly lead to higher success rates. 
The figure also shows that a higher
number of trees results in better performance.
Learning rates are between $0.08$ and $0.64$ give good results.
It is also apparent that deeper trees (at least 8) increase the accuracy.

\subsubsection{Experimental Setup}
The XGBoost model is evaluated on the task of tactic prediction both in the split setting and the chronological setting (illustrated in Section~\ref{sec:eval}).
We use the strong negative examples and determine the final parameters---$\textit{ratio}=16$, $\textit{eta}= 0.2$,
$\textit{number of trees} = 1024$, $\textit{max depth}=10$---for generating a model from non-sink nodes and use that to predict 
for sink nodes.

Since the entire dataset contains approximately $250,000$ proof states, and it is time-consuming to generate a unique XGBoost model for each test case,
we propose several ways to speed up the chronological evaluation.
Instead of training on the data from all preceding states, we merely provide $1,000$ instances occurring previously as the training data.
According to the results of parameter tuning depicted in Figure~\ref{fig:XGB-param-tuning}, we decide on the hyperparameters---$\textit{ratio}=4$, $\textit{eta}= 0.2$, $\textit{number of trees}= 256$, $\textit{max depth}=10$---to balance the accuracy and efficiency.

\section{Experimental Evaluation}
\label{sec:eval}
To compare the performance of the described machine learning models, we
perform three kinds of experiments: \textit{split} evaluation,
\textit{chronological} evaluation, and evaluation in Tactician. Achieving good
performance in the last type of evaluation is the main goal. All three machine
learning models are evaluated in the first two kinds of experiments, while in
Tactician we only evaluate $k$-NN and online random forest. This is because the
XGBoost system, while being potentially the strongest machine learner among
tested, may not be easily turned into an online learner and integrated into
Tactician. We adopt the original features---term and term pairs---for evaluation outside Tactician, whereas both the original features and the new
are tested on Tactician's benchmark.
To determine the relative importance of the feature classes described
in~\Cref{sec:features}, we benchmark the addition of each class separately in
Tactician. All evaluations are performed on data extracted from the standard
library of Coq 8.11.
\subsection{Split Evaluation}
\label{sec:split}

\begin{table}[tb!]
\caption{\label{tab:eval}
Performance of the three tested machine learning models in two types of
evaluation: using a split of the dataset and a chronological evaluation through
the dataset. top-$n$ refers to the frequency of
the correct tactic being present in the first $n$ predictions from a machine
learning model.
}
\centering
\begin{tabular}{l@{\hspace{0.5cm}}cc@{\hspace{0.3cm}}cc@{\hspace{0.5cm}}cc@{\hspace{0.2cm}}}
\toprule
&\multicolumn{6}{c}{\textbf{Machine learning system}} \\
\cmidrule[0.05em](r){2-7}
&\multicolumn{2}{c}{$k$-NN}
&\multicolumn{2}{c}{Random Forest}
&\multicolumn{2}{c}{XGBoost} \\
\cmidrule[0.05em](r){2-3}
\cmidrule[0.05em](r){4-5}
\cmidrule[0.05em](r){6-7}
\textbf{Evaluation type}
& top-1 & top-10
& top-1 & top-10
& top-1 & top-10 \\
\cmidrule[0.05em](r){1-1}
\cmidrule[0.05em](r){2-3}
\cmidrule[0.05em](r){4-5}
\cmidrule[0.05em](r){6-7}
split  	          & 18.8\% & 34.2\% & 32.1\% & 41.2\% & 18.2\% & 38.2\% \\
chronological     & 17.3\% & 43.7\% & 29.9\% & 58.9\% & 18.2\% & 43.4\% \\
\bottomrule
\end{tabular}
\end{table}

In the directed acyclic graph of dependencies of the Coq modules, there are 545
nodes. 104 of them are \textit{sink nodes}, i.e., these are the modules that do
not appear among dependencies of any other module. We used these modules as
final testing data for evaluation outside Tactician. The rest of the data was
randomly split into training and validation parts and was used for parameter
tuning of random forest and gradient boosted trees. The models with tuned
hyperparameters were evaluated on the testing data. The results of the evaluation of
the three tested models are shown in the first row of Table \ref{tab:eval}.

\subsection{Chronological Evaluation}
\label{sec:chrono}
Although the split evaluation from the previous experiment is interesting,
it does not correspond entirely to the Tactician's internal mode of
operation.
To simulate the real-world scenario in an offline setting, we create an individual model for each proof state by learning from all the previous
states---data from dependent files and preceding lines in the local file.
The second row of Table~\ref{tab:eval} presents the results of the evaluation in chronological order.

\subsection{Evaluation in Tactician}

\begin{table}[tb!]
\caption{\label{tab:tactician}
Proving performance of two online learners integrated with Tactician, $k$-NN and
random forest, in the Coq Standard Library.  The percentages in the
table correspond to the fraction of lemmas proved in a given Coq module. The columns \textit{union} show what fraction of the
lemmas was proved by at least one of the learners. RF is an
abbreviation of random forest.
}
\centering
\footnotesize
\begin{tabular}{l@{\hspace{0.2cm}}c@{\hspace{0.2cm}}ccc@{\hspace{0.2cm}}ccc@{\hspace{0.2cm}}}
\toprule
\textbf{Coq module} & \textbf{\#Lemmas} & \multicolumn{6}{c}{\textbf{Features type}} \\
\cmidrule[0.05em](r){1-1}
\cmidrule[0.05em](r){2-2}
\cmidrule[0.05em](r){3-8}
&&
\multicolumn{3}{c}{{Original}} &
\multicolumn{3}{c}{{New}} \\
\cmidrule[0.05em](r){3-5}
\cmidrule[0.05em](r){6-8}
&&
 $k$-NN & RF & \textit{union} &
 $k$-NN & RF & \textit{union} \\
\cmidrule[0.05em](r){3-5}
\cmidrule[0.05em](r){6-8}

\textbf{All}&\textbf{1137}&\textbf{33.7\%}&\textbf{35.3\%}&\textbf{39.6\%}&\textbf{34.7\%}&\textbf{36.2\%}&\textbf{40.4\%}\\
Arith&293&52\%&59\%&65\%&56\%&59\%&66\%\\
Bool&130&93\%&87\%&93\%&92\%&88\%&92\%\\
Classes&191&80\%&76\%&81\%&79\%&79\%&83\%\\
FSets&1137&32\%&34\%&37\%&32\%&35\%&39\%\\
Floats&5&20\%&20\%&20\%&40\%&19\%&40\%\\
Init&164&73\%&51\%&73\%&73\%&56\%&73\%\\
Lists&388&38\%&43\%&47\%&38\%&44\%&49\%\\
Logic&341&31\%&27\%&34\%&32\%&31\%&35\%\\
MSets&830&38\%&40\%&43\%&36\%&40\%&43\%\\
NArith&288&37\%&43\%&44\%&35\%&42\%&47\%\\
Numbers&2198&23\%&22\%&27\%&24\%&23\%&27\%\\
PArith&280&31\%&40\%&44\%&35\%&39\%&45\%\\
Program&28&75\%&64\%&75\%&78\%&66\%&78\%\\
QArith&295&33\%&40\%&43\%&31\%&39\%&45\%\\
Reals&1756&19\%&23\%&25\%&21\%&24\%&26\%\\
Relations&37&29\%&24\%&40\%&27\%&26\%&29\%\\
Setoids&4&1.00&1.00&1.00&1.00&97\%&1.00\\
Sets&222&43\%&42\%&49\%&49\%&47\%&53\%\\
Sorting&136&26\%&29\%&33\%&25\%&30\%&33\%\\
Strings&74&22\%&22\%&27\%&17\%&14\%&20\%\\
Structures&390&45\%&49\%&54\%&51\%&51\%&56\%\\
Vectors&37&37\%&29\%&40\%&21\%&23\%&27\%\\
Wellfounded&36&19\%&05\%&19\%&16\%&13\%&16\%\\
ZArith&953&41\%&46\%&49\%&40\%&43\%&46\%\\
btauto&44&11\%&20\%&20\%&20\%&17\%&22\%\\
funind&4&75\%&50\%&75\%&50\%&73\%&75\%\\
micromega&339&21\%&27\%&29\%&27\%&25\%&30\%\\
nsatz&27&33\%&33\%&37\%&40\%&26\%&40\%\\
omega&37&40\%&67\%&67\%&48\%&63\%&64\%\\
rtauto&33&30\%&39\%&48\%&33\%&44\%&51\%\\
setoid\_ring&362&21\%&23\%&26\%&27\%&27\%&30\%\\
ssr&311&68\%&55\%&69\%&70\%&57\%&71\%\\
\bottomrule
\end{tabular}
\end{table}

Table \ref{tab:tactician} shows the results of the evaluation of two online learners---the $k$-NN and the random forest---within Tactician. The hyperparameters of
the random forest model were chosen based on the grid search in \Cref{sec:rf-tuning}.
We run the proof search for every lemma in the library with a 40-second time
limit on both the original and the improved features.

The random forest performed marginally better than $k$-NN on both kinds of features.
With old features the $k$-NN proved 3831 lemmas (being 33.7\% out of
all 11370), whereas the random forest proved 4011 lemmas (35.3\% of all). With the
new features, both models performed better, and again, the random forest proved
more lemmas (4117, 36.2\% of all) than $k$-NN (3945, 34.7\% of all).

It is somewhat surprising that the random forest, which performed much better
than $k$-NN on the split in the offline evaluation, is only better by a small
margin in Tactician. This may be related to the time and memory consumption of
random forest, which may be higher than for $k$-NN on certain kinds of
data.\footnote{Doing the splits in the leaves has quadratic time complexity
with respect to the number of examples stored in the leaf; sometimes it happens,
that leaves of the trees store large number of examples.}

It is worth noting that $k$-NN and random forest resulted in quite
different sets of proofs. The columns marked as \textit{union} show that the
size of the union of proofs constructed by the two models is significantly larger
than the number of proofs found by each model separately. In total, both
models resulted in 4503 (39.6\%) proofs using old features and 4597 (40.4\%)
proofs using the new features.

\subsection{Feature Evaluation}
Table \ref{tab:feature} depicts the influence of adding the new classes of features
described in \Cref{sec:features} to the previous baseline.\footnote{The results here are not directly comparable to those in Table \ref{tab:tactician} mainly due to the usage of a non-indexed version of $k$-NN in contrast to the algorithm presented in \ref{alg:lshf}.}
All of the newly produced features improve the success rates.
However, the top-down oriented AST walks contribute little, probably due to Tactician having already included term tree walks up to length $2$.
Every other modification obtains a reasonable improvement, which confirms the
intuitions described in Section \ref{sec:features}.

\begin{table}[tb!]
	\caption{\label{tab:feature}
		Proving performance of each feature modification.
		$\mathcal{O}, \mathcal{W}, \mathcal{V}, \mathcal{T}, \mathcal{S}, \mathcal{C}$ denote original features, top-down oriented AST walks, vertical abstract walks, top-level structures, premise and goal separation, and adding feature occurrence, respectively. The symbol $\oplus$ denotes that we combine the original features and a new modification during the experiments.
	}
	\centering
	\setlength\tabcolsep{4pt}

	\begin{tabular}{lcccccc}
		\toprule
		Features
		& $\mathcal{O}$
		& $\mathcal{O} \oplus \mathcal{W}$
		& $\mathcal{O} \oplus \mathcal{V}$
		& $\mathcal{O} \oplus \mathcal{T}$
		& $\mathcal{O} \oplus \mathcal{S}$
		& $\mathcal{O} \oplus \mathcal{C}$
		\\

		\hline
		Success rates ($\%$)
		& 32.75 & 32.82
		& 34.16 & 33.65
		& 34.42 & 34.97 \\
		\bottomrule
	\end{tabular}
\end{table}

\section{Related Work}\label{sec:related-work}

Random forests were first used in the context of theorem proving by Färber~\cite{mfck-frocos15},
where multi-path querying of a random forest would improve on $k$-NN results for premise
selection. Nagashima and He~\cite{NagashimaH18} proposed a proof method recommendation
system for Isabelle/HOL based on decision trees on top of precisely engineered
features. A small number of trees and features allowed for explainable recommendations.
Frameworks based on random boosted trees (XGBoost, LightGBM) have also been used in automated
reasoning, in the context of guiding tableaux connection proof search~\cite{ckjuhmmo-nips18}
and the superposition calculus proof search \cite{ChvalovskyJ0U19}, as well as for handling
negative examples in premise selection \cite{PiotrowskiU18}.

Machine learning to predict tactics was first considered by Gauthier et al.~\cite{tgckju-lpar17}
in the context of the HOL4 theorem prover. His later improvements \cite{tgckjurkmn-jar21} added
Monte-Carlo tree search, tactic orthogonalization, and integration of both Metis and a hammer
\cite{tgck-cpp15}. A similar system for HOL Light was developed by Bansal et al.
\cite{BansalLRSW19}. Nagashima and Kumar developed the proof search
component~\cite{NagashimaK17} of such a system for Isabelle/HOL. This work builds upon
Tactician~\cite{DBLP:conf/mkm/BlaauwbroekUG20,blaauwbroek2020tactic}, adapting and improving
these works for dependent type theory and the Coq proof assistant.

\section{Conclusion}

We have implemented several new methods for learning tactical guidance
of Coq proofs in the Tactician system.  This includes better proof
state features and an improved version of approximate $k$-nearest
neighbor based on locality sensitive hashing forests. A completely new
addition is our online implementation of random forest in Coq, which
can now be used instead of or together with the $k$-nearest neighbor.  We
have also started to experiment with strong state-of-the-art learners
based on gradient boosted trees, so far in an offline setting using binary learning with negative examples.

Our random forest improves very significantly on the $k$-nearest
neighbor in an offline accuracy-based evaluation. In an online
theorem-proving evaluation, the improvement is not as big, possibly
due to the speed of the two methods and the importance of backtracking
during the proof search. The methods are, however, quite complementary
and running both of them in parallel increases the overall performance of
Tactician from 33.7\% ($k$-NN with the old features) to 40.4\% in 40s.
Our best new method (RF with the new features) now solves 36.2\% of the problems in 40s.

The offline experiments with gradient boosted trees are so far
inconclusive. They outperform $k$-nearest neighbor in top-10 accuracy,
but the difference is small, and the random forest performs much better
in this metric. Since the random forest learns only from positive
examples, this likely shows that learning in the binary setting with
negative examples is challenging on our Tactician data. In particular,
we likely need good semantic feature characterizations of the tactics,
obtained e.g., by computing the difference between the features of the
proof states before and after the tactic application. The experiments,
however, already confirm the importance of choosing good negative data to
learn from in the binary setting.

\end{document}